\documentclass[reprint,amsmath,amssymb,aps,prl]{revtex4-1}

\usepackage{graphicx}
\usepackage{dcolumn}
\usepackage{bm}
\usepackage{xcolor}
\usepackage{verbatim}
\usepackage{soul}

\begin{document}

\preprint{APS/123-QED}

\title{Bipolar thermoelectric superconducting single-electron transistor}

\author{Sebastiano Battisti}
\email{sebastiano.battisti@sns.it}
\affiliation{NEST, Istituto Nanoscienze-CNR and Scuola Normale Superiore, I-56127 Pisa, Italy}

\author{Giorgio De Simoni}
\affiliation{NEST, Istituto Nanoscienze-CNR and Scuola Normale Superiore, I-56127 Pisa, Italy}

\author{Luca Chirolli}
\affiliation{NEST, Istituto Nanoscienze-CNR and Scuola Normale Superiore, I-56127 Pisa, Italy}

\author{Alessandro Braggio}
\email{alessandro.braggio@nano.cnr.it}
\affiliation{NEST, Istituto Nanoscienze-CNR and Scuola Normale Superiore, I-56127 Pisa, Italy}

\author{Francesco Giazotto}
\email{francesco.giazotto@sns.it}
\affiliation{NEST, Istituto Nanoscienze-CNR and Scuola Normale Superiore, I-56127 Pisa, Italy}

%\date{\today}

\begin{abstract}

Thermoelectric effects in normal metals and superconductors are usually very small due to the presence of electron-hole symmetry. Here, we show that superconducting junctions brought out of equilibrium manifest a sizable bipolar thermoelectric effect that stems from a \emph{strong} violation of the detailed balance. To fully control the effect, we consider a thermally biased \textit{SIS'IS} junction where the capacitance of the central \textit{S'} region is small enough to establish a Coulomb blockade regime. By exploiting charging effects we are able to tune the Seebeck voltage, the thermocurrent, and thereby the power output of this structure, via an external gate. We then analyse the main figures of merit of bipolar thermoelectricity and we prospect for possible applications.
\end{abstract}

\maketitle

\textit{Introduction\textemdash}
Thermal transport and quantum thermodynamics at the nanoscale have recently attracted a growing interest \cite{benenti2017,dubi2011,kosloff2013,seifert2012,campisi2011,giazottoOpportunities2006,fornieri2017,muhonen2012,brunner2102,barato2015,polettini2015,verley2014,pietzonka2018,manikandan2019,pekola2007radio,saira2007heat,fornieri2014normal,tirelli2008manipulation,sothmann2017high}, thanks to the opportunity the thermoelectric effect offers to manipulate heat and control the energy efficiency of
nanodevices  \cite{claughton1996,ozaeta2014,vischi2019,giazotto2015a,giazotto2014,marchegiani2016,brandner2013,sothmann2014,esposito2009,whitney2014,ronetti2016,kamp2019,hussein2019}. In the linear regime thermoelectricity requires a broken electron-hole (EH) symmetry, which also implies  a non-reciprocal $IV$ characteristic, i.e., $I(V,\Delta T)\neq -I(-V,\Delta T)$, where $\Delta T$ is the temperature difference. Indeed metals, that are almost electron-hole (EH) symmetric, show nearly negligible Seebeck coefficients \cite{mott1958} and  present zero thermovoltages in the superconducting phase. However, in the non-linear regime it has been demonstrated \cite{marchegiani2020,marchegiani2020a}, that superconducting \textit{SIS'} tunnel  junctions with a sufficiently suppressed Josephson coupling exhibit a sizable thermopower due to the spontaneous breaking of EH symmetry, yielding an effective Seebeck coefficient (${\cal S}$) as large as  $\sim 10^5$ times its value in the normal state \cite{germanese2022,germanese2023phase}. At the same time, the EH symmetry determines the full \emph{bipolarity} of the effect with reciprocal $IV$ characteristics. The bipolar thermoelectric effect emerges when a strong temperature difference is suitably applied, and in the presence of strong asymmetry in the energy gaps of the junction.

In this Letter, we consider an \textit{SIS'IS} structure where a central superconducting (SC) island  featuring strong Coulomb interaction is sandwiched between two SC leads via tunnel barriers. In such a system, the origin of the bipolar thermoelectric properties greatly differs from that of standard thermoelectricity in quantum dots, and lies in the \emph{strong} violation of the detail balance induced by the temperature difference in the junction and the interacting nature of BCS theory. Furthermore, we exploit the gating properties of the Coulombic island to control the thermoelectric performances of the engine.  This unique electrical tunability  differs from other platforms \cite{bernazzani2023}, and can be  relevant for on-chip energy harvesting and other energy management purposes in superconducting quantum processors and radiation sensors \cite{paolucci2023highly}.\\

\begin{figure}[t!]
  \includegraphics[width=1.1\columnwidth,trim= 1.6cm 0cm 0cm 0cm]{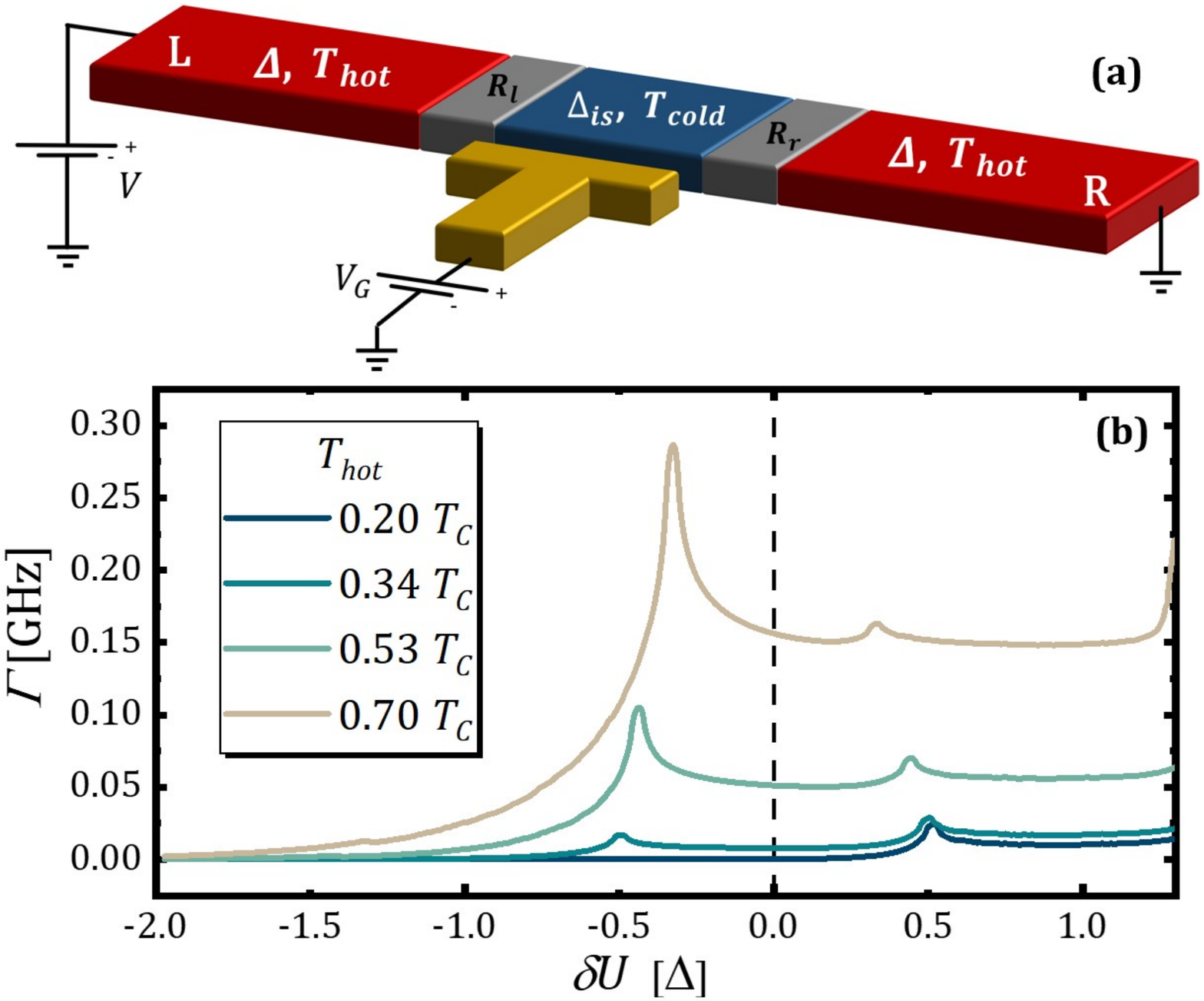}
    \caption{
    \textbf{(a)}: Scheme of the \textit{SIS'IS} transistor. The red-coloured parts show the hot superconductors and the blue-coloured part shows the cold one. $V$ and $V_G$ denote respectively the source-drain and gate voltages.  \textbf{(b)}: Sequential tunneling rates versus energy for different values of $T_{hot}$ at $T_{cold} = 0.2~  T_{C}$. Here we set $\Delta=220~\mu$eV (corresponding to Al, aluminum), $\Delta_{is}=\Delta/2$, which can be obtained using a normal metal-superconducting bilayer \cite{germanese2022,hijano2023}, and all identical barrier resistances $R=1{\rm M}\Omega$.}
  \label{fig1}
\end{figure}

\begin{figure*}[ht!]
  \includegraphics[width=2\columnwidth,trim=2cm 10cm 2cm 12cm]{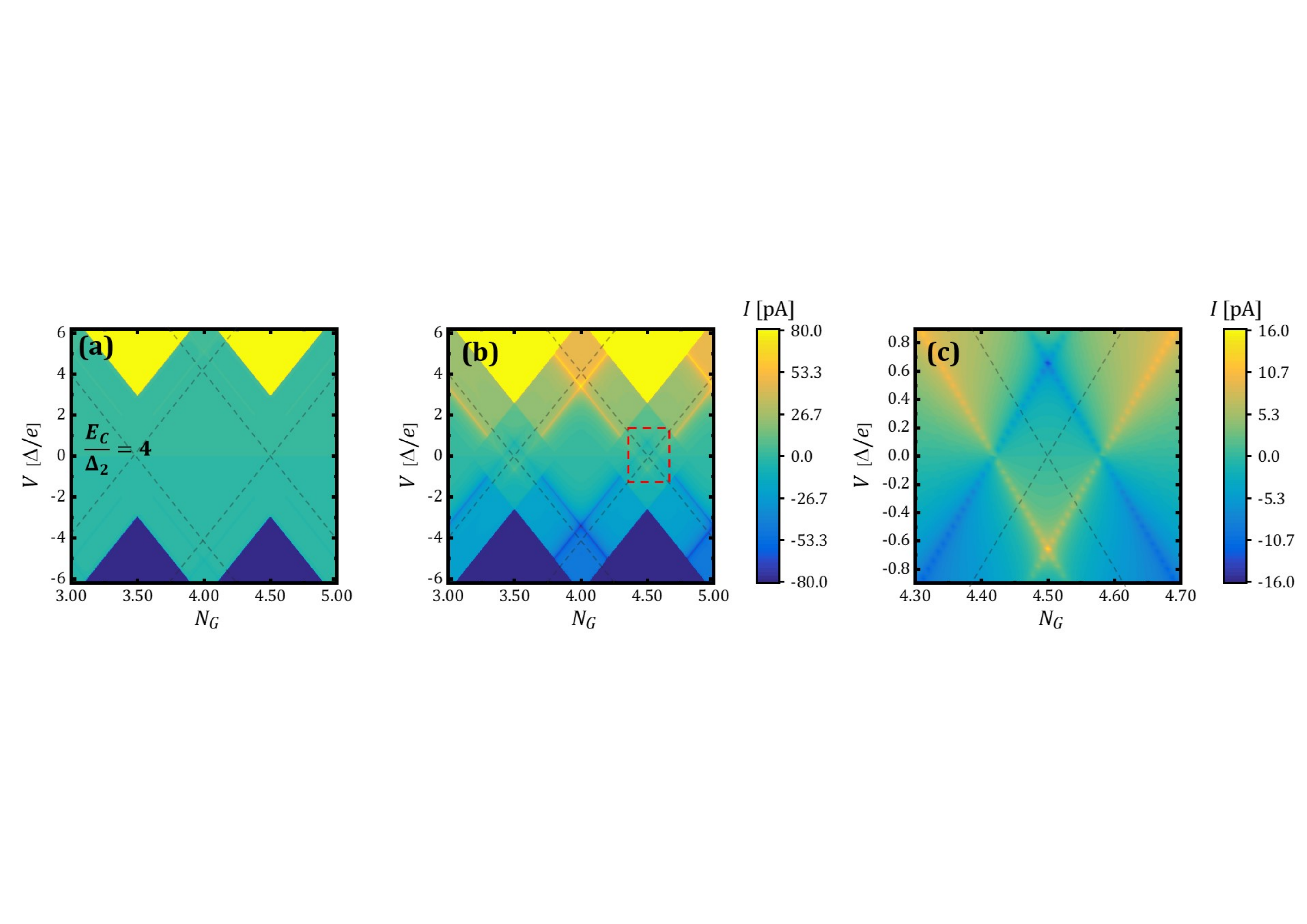}
    \caption{\textbf{(a)},\textbf{(b)}: Charge current flowing through the \textit{SIS'IS} structure at thermal equilibrium (\textbf{a}) and in the non-linear regime (\textbf{b}) versus $V$ and $N_G=C_G V_G /e$. The colour scale is the same for both graphs. \textbf{(c)}: Blow-up of the red-dashed region shown in panel (\textbf{b}) where the onset of the thermoelectric behaviour is clearly visible. The calculations parameters are as in Fig.~\ref{fig1}.}
  \label{fig2}
\end{figure*}

\textit{Model\textemdash}The \textit{SIS'IS} structure under investigation is shown in Fig.\ref{fig1}(a), and  consists of two superconducting (SC) leads (\textit{L,R}, red part in Fig.\ref{fig1}(a)) with SC gap $\Delta$ put in tunnel contact with a Coulombic island (central blue part in Fig.\ref{fig1}(a)) with a different SC gap $\Delta_{is}$, via two identical barriers of resistance $R_{L/R}$.  In order to observe bipolar  thermoelectricity the leads are chosen to have a larger gap than the island, $\Delta > \Delta_{is}$, and they are kept at a temperature $T_{hot} > T_{cold}$ larger than  the island temperature, $T_{is}\equiv T_{cold}=0.2~ T_{C}$, where $T_{C}$ is the critical temperature of the superconducting  leads. The tunneling barriers are assumed to be resistive enough to make the Josephson energy negligible with respect to thermal energy, thus allowing  the Josephson coupling to be neglected \cite{marchegiani2020c}. Yet, in order to observe Coulomb blockade we assume the charging energy of the central island, $E_C=e^2/2C_{tot}$, with $C_{tot}$ the total island capacitance, large enough that $E_C \gg k_B T_l$ with $l=is,L,R$ and $E_C\gtrsim \Delta$.

For sufficiently resistive barriers the full transport properties of the system can be described through the rates $\Gamma_j(\delta U)$, with $j=R,L$, that describe the tunneling probability through the $j$-th barrier by the Fermi golden rule \cite{averin1991,pekola2008}
\begin{equation}
   \Gamma_j= \frac{1}{e^2 R_j} \int_{-\infty}^\infty
   \!\!\!\!\!\!\! dE ~ n_{is}(E + \delta U)[1-f_{is}(E+\delta U)]
    n_j(E) f_j(E),
    \label{Eq:rates} 
\end{equation}
where $e$ is the electron charge, $\delta U$ is the electrostatic energy acquired ($\delta U>0$) or lost ($\delta U<0$) during the tunneling process \cite{averin1991},  $n_{l}(E)=\left| Re[(E+i\gamma)/\sqrt{(E + i\gamma)^2 - \Delta^2_l (T_l)}]\right|$ is the (smeared by non-zero $\gamma \ll \Delta_l$) BCS density of states (DOS) of the $l=is,L,R$ element  \cite{dynes1984}, and $f_{l}(E)$ the Fermi-Dirac distribution function at equilibrium temperature $T_l$. We assumed the SC island to be fully in equilibrium at its temperature $T_{is}$.\\

\textit{Strong violation of detail balance\textemdash}The electron tunneling rate from the lead to the island as a function of the electrostatic energy difference $\delta U$ contains the crucial physics for the bipolar thermoelectric effect, and it is shown in Fig.~\ref{fig1}(b) for different temperatures $T_{hot}$. 
At equilibrium,  $T_{hot}=T_{cold}$ (dark blue line), we notice a small peak at $\delta U^*=\Delta - \Delta_{is}>0$, which corresponds to the matching of the DOS divergences of the two different  superconductors and it is activated by the temperature, since at zero temperature it is completely Pauli blocked. At finite temperature in equilibrium, the rates satisfy the detail balance $\Gamma(-\delta U)=e^{-\delta U/k_B T}\Gamma(\delta U)$, so correspondingly we expect a small peak [although not visible in Fig.~\ref{fig1}(b)] to exist at negative energies $\delta U=-\delta U^*$ as well .

More intriguing physics arises in out-of-equilibrium conditions. By increasing the lead  temperature, $T_{hot}> T_{cold}$, the peak at negative energy emerges and becomes much stronger than its counterpart at positive energy (brown line). This is an unusual situation where  $\Gamma(-|\delta U|)>\Gamma(|\delta U|)$, and we identify it as a {\it strong violation of the detail balance}. 
It comes out as a consequence of a finite temperature difference that populates more states at higher energy in the hot leads. From
Eq.~(\ref{Eq:rates}) a strong violation is possible \emph{only if} for some energy $\epsilon, \delta U >0$ we have
\begin{equation}
\label{eq:inequality}
n_{is}(\epsilon-\delta U)[1-f_{is}(\epsilon-\delta U)]>n_{is}(\epsilon+\delta U)[1-f_{is}(\epsilon+\delta U)]. 
\end{equation}
This inequality constitutes a {\it necessary} condition and it depends only on the island DOS and on its Fermi function.
It can be shown 
\footnote{For more information refer to the Supplementary Material.}
that Eq.~(\ref{eq:inequality}) is meaningful \emph{only if}
there is a gap asymmetry between the hot and cold side,  and the superconducting DOS features a monotonously decreasing energy dependence for $|E|>\Delta$ \cite{marchegiani2020}. Furthermore, in order to be satisfied, Eq.~(\ref{eq:inequality}) \emph{implicitly} requires that the BCS island DOS shifts with the electrochemical potential (the electrostatic energy difference $\delta U$ in the rate). Indeed, if the DOS does not depend on $\delta U$ (i.e., substituting $n_{is}(\epsilon\pm\delta U)\to n_{is}(\epsilon)$ like in a semiconductor) 
Eq.~(\ref{eq:inequality}) cannot be satisfied since $n_{is}>0$. 
Thus, the strong violation stems essentially from the interacting character captured by the BCS mean field theory and, as we are going to show,it is the essential precursor of the bipolar thermoelectric phenomena. The fundamental role of the interaction clarifies also well why this effect can be associated to a spontaneous symmetry breaking of the PH symmetry induced by the out-of-equilibrium condition \cite{marchegiani2020,germanese2022,germanese2023phase}.\\

{\it Current\textemdash}In the sequential tunneling approximation we get the electric current through a standard master equation approach \cite{nazarovbook2009,bagrets2004,braggio2006,flindt2008}, in which the time-evolution of the population $P_n$ of the island charge state $n$ is determined by $\dot{P}_{n}=\sum_{n'}\mathbf{W}_{nn'}P_{n'}$. The kernel $\mathbf{W}_{nn'}=\sum_{j=R, L}\Gamma_j^{nn'}$ with the diagonal terms $\mathbf{W}_{nn}$ fixed by the conservation of the probability, 
$\sum_n P_n=1$ implying $\sum_n\mathbf{W}_{nn'}=0$. The rates $\Gamma_j^{n,n\pm 1}=\Gamma_j(\delta U_{n, n\pm1})$ correspond only to the transitions $n= n'\pm 1$ described by the rates of Eq.~(\ref{Eq:rates}). The electrostatic energy difference is $\delta U_{n, n\pm 1}=U(n\pm 1)-U(n)\pm e V/2$ \cite{averin1991,beenakker1991,nazarovbook2009}, with $V=V_L-V_R$ the source-drain bias, and $U(n)=E_C (n-q_{is}/e)^2/2$ the electrostatic energy, which depends on the offset charge on the island $q_{is}=C_g V_g+\sum_j C_j  V_j$,  with $C_g$ the gate capacitance and $C_j$ the $j$-th barrier capacitance (for which $C_{tot}= C_g+C_L+C_R$). In the stationary limit \cite{beenakker1991}, the current can be simply computed in the right lead  $I=I_R=-I_L=e \sum_n [\Gamma_R^{n, n+1}-\Gamma_R^{n, n-1}]P_n^0$ where $P_n^0$ is the stationary probability of island charge states. This general numerical approach can be further simplified in the Coulomb blockade regime $k_B T\ll E_C$ by noting that the dominant contribution to the transport at one resonance is associated only to the tunneling rates involving neighboring charge states $n-1\rightleftharpoons n$.  The current in such case can be written as 
\begin{equation}\label{Eq:Isimple}
   I_{n}=e\frac{\Gamma_{L}^f(n-1) \Gamma_{R}^f(n)-\Gamma_{L}^b(n) \Gamma_{R}^b(n-1)}{\Gamma_{L}^f(n-1)+\Gamma_{R}^f(n)+\Gamma_{L}^b(n) +\Gamma_{R}^b(n-1)},
   \end{equation}
where we use a shortened notation for the rates $\Gamma_{j}^{f/b}(n)=\Gamma_j(\pm\delta U_{n, n\pm 1 })$. Since we are mainly interested in the deep subgap regime in the bias range $|eV|<2\Delta$ with $E_C=4\Delta$,  we  approximate the current by considering the dominant contribution of two neighboring Coulomb resonances, i.e., $I=I_n+I_{n+1}+\mathcal{O}(e^{-E_c/2k_BT_{hot}})$. \\
\begin{figure}[t]
\includegraphics[width=1.15\columnwidth,trim=3cm 0cm -0.5cm 0cm]{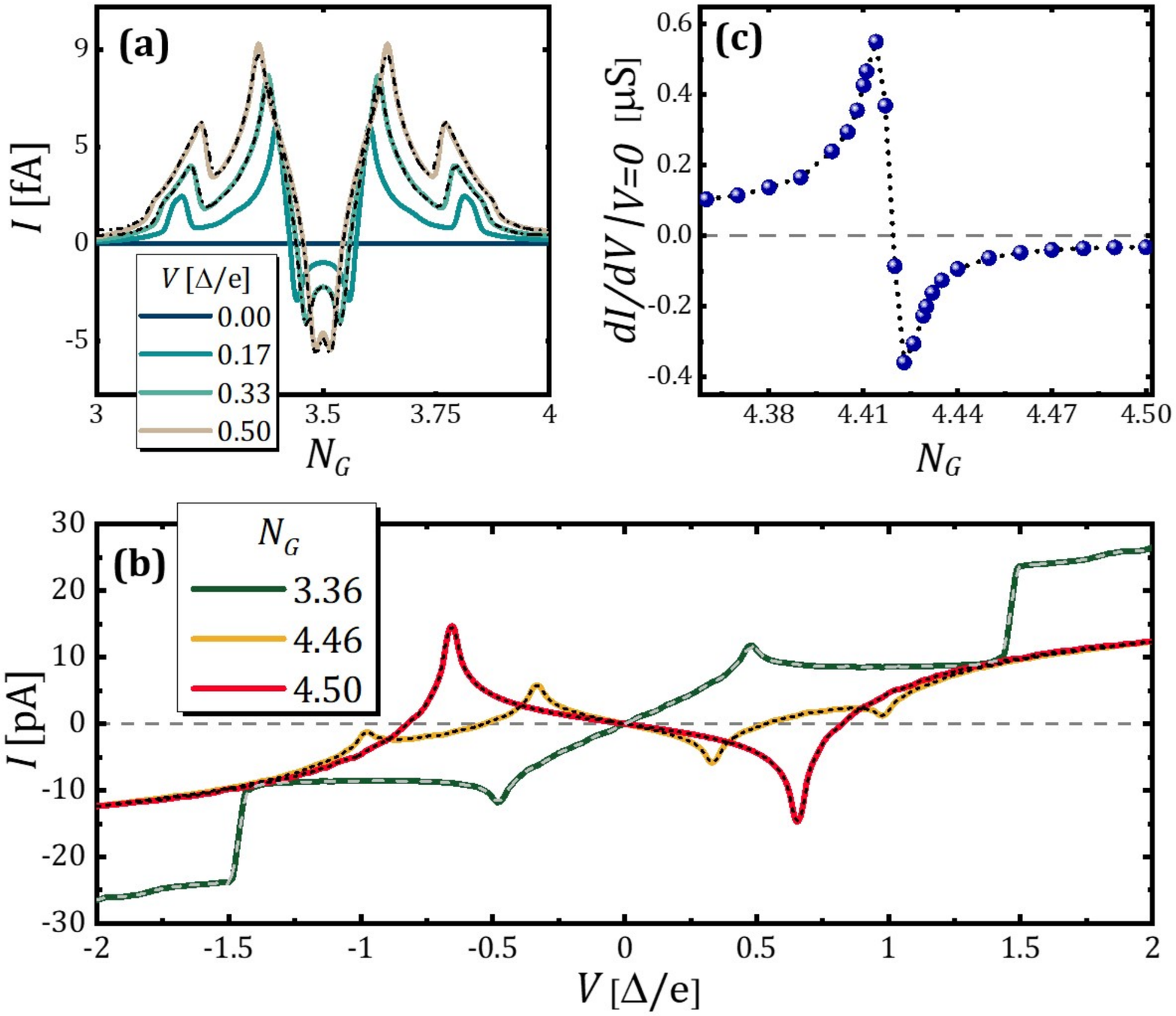}
    \caption{Out-of-equilibrium ($T_{hot}=0.7 T_{C}$) current $I$ as a function of: (\textbf{a}) $N_G$ for different values of $V$, and (\textbf{b}) $V$ for different values of $N_{G}$ (respectively horizontal and vertical cuts of Fig.~\ref{fig2}(\textbf{b})). The dashed lines correspond to the prediction of the simplified model Eq.~(\ref{Eq:Isimple}), showing its validity in the case of $E_C\gg e V$. \textbf{(c)}: Out-of-equilibrium zero-bias conductance as a function of $N_G$ calculated for the curves in (\textbf{b}).}
  \label{fig3}
\end{figure}

\textit{Coulomb diamonds\textemdash} Figure~\ref{fig2}(a),(b)  show the current $I$ as a function of the source-drain bias $V$ and the gate-tunable offset charge $N_G=C_g V_g/e$. Typical Coulomb diamonds appear, which display periodicity in the offset charge $N_G$ in units of the electron charge $e$. The system does not present any even-odd effect since the average Cooper pair recombination rate in our system  $\Gamma_r\simeq 16 $kHz \cite{maisi2013} is much smaller than the tunneling rates $\Gamma_j\sim I/e$ (inverse average electron dwelling time in the island). Coulomb diamonds at equilibrium are shown in Fig.~\ref{fig2}(a). 
As a guide for the eye, we show with black dashed lines the boundaries of the Coulomb diamonds for $\Delta=0$, where the electrostatic energies vanish, $\delta U_{n, n\pm 1}(N_G,V)\equiv 0$. 
As expected, the SC gap pushes the boundaries of the Coulomb diamonds up in energy, and charge transport is suppressed in the $(N_G,V)$ plane domains satisfying ($eV < 2\Delta + 2\Delta_{is}\approx 3\Delta$). 

In Fig.~\ref{fig2}(b) we show the results for the non-equilibrium case, $T_{hot}  =0.7~ T_C >T_{cold}$. 
Subgap conduction channels become clearly visible, as thermally excited states promote the stronger emergence of the matching peak resonances. 
At integer $N_G$ and $|eV|\sim 3\Delta$, where the transport is dissipative ($IV>0$) even if fully inside the Coulomb blockade diamond, we clearly see the appearance of yellow (blue) crosses at positive (negative) $V$. 
These features stems from the  enhancement of the \emph{negative energy} peak in the tunneling rate for  electrostatic energy $\delta U\approx -\delta U^*$ [see Fig.~\ref{fig1}(b)], and are also a direct consequence of the strong violation of the detailed balance, even if their nature is still dissipative. 
More intriguingly, for half-integer $N_G$ and $eV\sim \Delta/2$, the sign of the current becomes opposite to the bias $IV <0$, as shown in the blow-up of Fig.~\ref{fig2}(c). This behavior is a  signature of thermoelectricity and it appears \emph{only} when a finite temperature difference is applied between the island and the hot leads. These subgap structures are equivalently present at positive and negative bias for the \emph{same} temperature difference, showing the full \emph{bipolar} character that is enforced by the EH symmetry of the unbiased system. The emerging bipolar thermoelectric effect is similar to SIS' systems \cite{marchegiani2020b}. 
Furthermore, our bipolar thermoelectric superconducting transistor offers the possibility to be  manipulated thanks to  Coulomb interaction and associated gating effects. 
We also notice that, unlike a conventional quantum-dot thermoelectric effect \cite{staring1993,mahan1996} that is activated by a temperature gradient \emph{between} the leads, the present effect appears when the leads have the \emph{same} temperature higher than the Coulomb island \cite{Note1}.

Figure ~\ref{fig3}(a) displays cuts of Fig.~\ref{fig2}(b) at different $V$. In a region around half-integer gate charge $N_G$ we clearly see a change of the sign of the current around the Coulomb resonance opposing to the bias (thermoelectric effect). 
Notably, the sign of the thermoelectric current does not change passing through the resonant value due to the unique bipolar nature, unlike in the conventional (unipolar) thermoelectric effect in a quantum dot \cite{sothmann2015}.

\begin{figure}[t!]
  \includegraphics[width=1.15\columnwidth,trim=3cm 0cm -0.5cm 0cm]{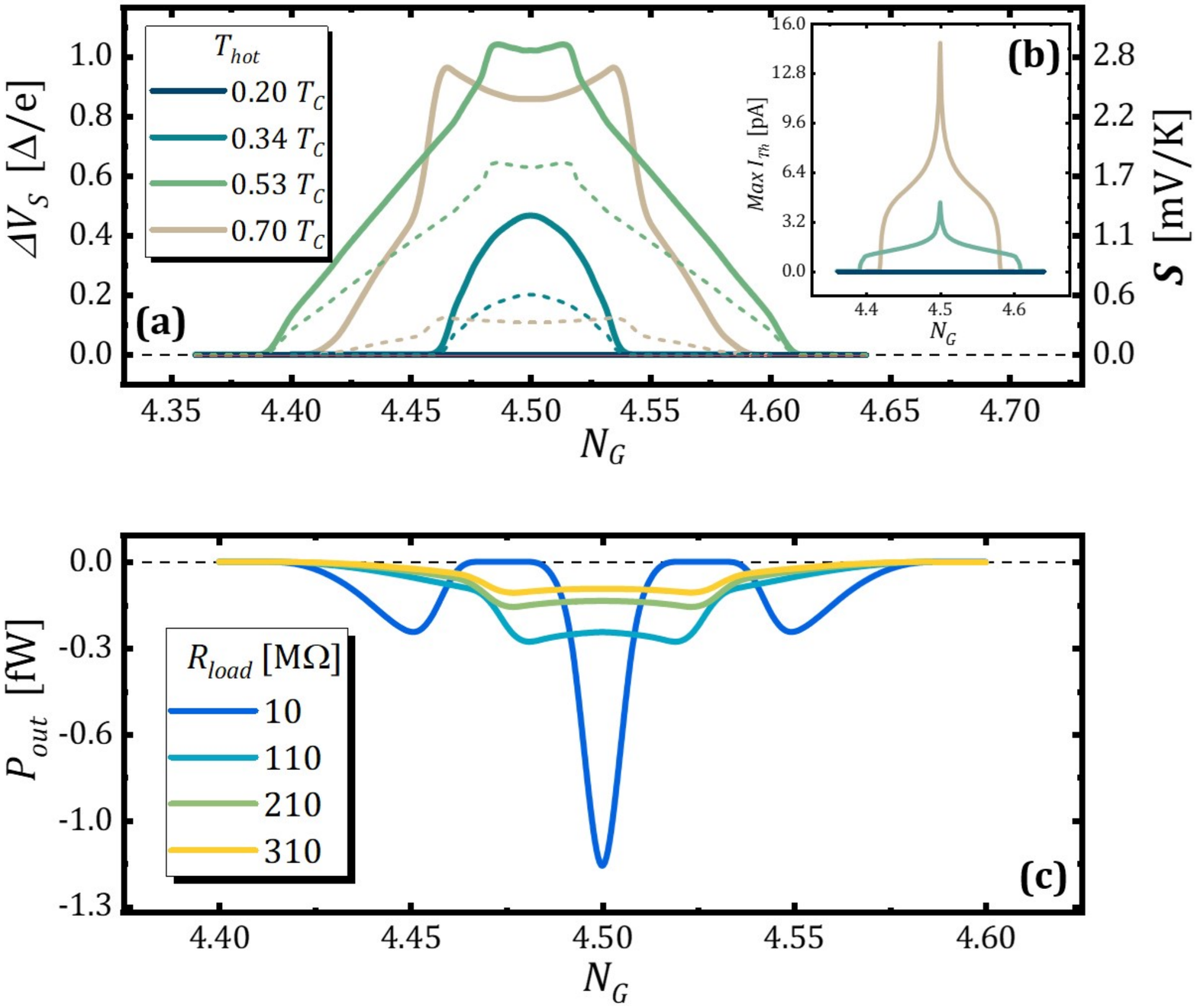}
    \caption{\textbf{a}: Seebeck voltage (left y-axis) and nonlinear Seebeck coefficient $|\mathcal{S}|$ (right y-axis) versus $N_G$ for different values of $T_{hot}$.
    \textbf{b}: Maximum thermocurrent versus $N_G$ for different values of $T_{hot}$.
    \textbf{c}: Power output $P_{out}=-\widetilde{I}~\widetilde{V}$ versus $N_G$ for different values of the load resistor. $\widetilde{I}$ and $\widetilde{V}$ are the solutions of the intersection of the $IV$ characteristics with the load line of the resistor. Note that among all  possible solutions only the electrically stable one, i.e., that with $dI/dV>0$, can be operated by the engine \cite{marchegiani2020,marchegiani2020c}.}
  \label{fig4}
\end{figure}
In Fig.~\ref{fig3}(b) we show cuts of Fig.~\ref{fig2}(b) at fixed $N_G$. 
We first focus on the  zero-bias behavior: as we vary $N_G$ towards half-integer values the zero-bias conductance (ZBC) $G_0=dI/dV |_{V=0}$ becomes negative at the critical value $N^*_G\approx 4.42$ [see Fig.~\ref{fig3}(c)]. 
This behavior blatantly suggests the spontaneous breaking of the EH symmetry of the system \cite{marchegiani2020,marchegiani2020b}, and highlights the unique capability of the gate control in our system that, differently from  other platforms \cite{bernazzani2023}, allows  to continuously tune the emergent bipolar thermoelectric properties.  A negative ZBC, together with the condition $I(0,\Delta T)=0$  as dictated by EH symmetry, implies the existence of thermoelectricity ($IV<0$) and the existence of a Seebeck voltage $V_S$, since at high-biases $eV\gg 3\Delta$ the system necessarily becomes again dissipative, $IV>0$. Interestingly, this implies that the Seebeck voltage (i.e., open circuit voltage defined as $I(V_S)=0$) is expected to be dependent on the gate voltage, $V_S(N_G)$, with clear consequences on the gate tunability of the thermoelectric performance.
At finite values of $V$ the current exhibits a peak changing in sign (thermoelectricity) when $N_G$ approaches half-integer values. The system shows an absolute negative conductance and thereby thermopower, $\dot{W}=-I V>0$.  
Furthermore, for some values of $N_G$ the $IV$ curve presents more than one resonant peak (see yellow line): this is a consequence of the Coulomb blockade,
since the island electrostatic energy differences $\delta U$ for different rates depends on the gate $N_G$ and bias $V$, and correspondingly the matching peaks resonance of the dominant rates appear split for non resonant values $N_G\neq 1/2 + n$ of the $IV$ curve. 
Note that a similar double-peak structure can be observed also as a function of $N_G$, as shown in Fig.~\ref{fig3}(b).\\

\textit{Thermoelectric figures of merit\textemdash}The intrinsic nonlinear nature of the above effect does not allow to describe the thermoelectric figures of merit of our system via a linear thermoelectric approach \cite{goldsmid2013}. However, we can still define a Seebeck voltage $V_S$ and a nonlinear Seebeck coefficient ${\cal S}=V_S/\Delta T $ with $\Delta T=T_{hot}-T_{cold}$. 
We stress that EH symmetry implies two Seeback voltages, $\pm V_S$, and a \emph{bipolar} ${\cal S}$~\cite{marchegiani2020}. Figure~\ref{fig4}(a)  shows $V_S$ (solid line, left scale) and $|{\cal S}|$ (dashed line, right scale) as a function of $N_G$ for different temperatures of the leads. By changing $T_{hot}$, the Seebeck voltage shows horn-like nonlinear features which are even higher at slightly lower value of $T_{hot}$. 
By inspection of Fig.~\ref{fig3}(b)  we recognize that the yellow curve has two peaks and for certain values of $\Delta T$ the second peak can even cross the $I=0$ axis, returning a higher open circuit (Seebeck) voltage. Analogously, ${\cal S}$ is also similarly affected and its maximal value is not necessarily associated to the maximal thermovoltage (due to the nonlinearities it is not even necessarily associated with the highest temperature difference $\Delta T$). In Fig.~\ref{fig4}(b) the maximum thermocurrent $I_{max}(N_G)=\operatorname{max}_{0<V<2\Delta/e}[|I(V,N_G)|]$ as a function of $N_G$  is shown to gradually become zero while lowering $T_{hot}$, as expected, since the temperature difference is not enough to trigger the bipolar thermoelectricity \cite{germanese2022,germanese2023phase}.
The thermoelectric generator character of the transistor appears when closing the circuit on a load resistor. 
Figure~\ref{fig4}(c) displays the output power  of the structure as a function of $N_G$ for different values of the load resistor, demonstrating the ability of fine gating control of the output power. The maximum achievable output power is typically associated with the smallest possible load resistance, and turns out to depend on several parameters.\\

\textit{Conclusions\textemdash}We theoretically proposed and analysed a bipolar thermoelectric superconducting single-electron transistor that enables tuning and control of the bipolar thermoelectric effect through an applied gate voltage. 
The interplay between Coulomb blockade and  out-of-equilibrium thermoelectricity finds its origin in the strong violation of the detail balance, which  
is triggered by different SC gaps, a finite temperature difference and, crucially, by the interacting nature intrinsic to the BCS theory. We investigated the performance of a fully gate-tunable heat engine that can provide, with realistic parameters, a nonlinear Seebeck coefficient up to $\sim 3$ mV/K at subKelvin temperatures. 
The effect can be implemented in a device that can produce gate-controlled single-electron thermoelectricity in a fully superconducting design, thereby fostering interest for on-chip energy harvesting and management, single-charge electronics, and single-photon detection \cite{paolucci2023highly}.\\

We  acknowledge the
EU’s Horizon 2020 Research and Innovation Framework
Programme under Grant No. 964398 (SUPERGATE),
No. 101057977 (SPECTRUM), and the PNRR MUR project PE0000023-NQSTI for partial financial support. A.B. acknowledges the Royal Society through the International Exchanges between the UK and Italy (Grants No. IEC R2 192166.).

\bibliographystyle{apsrev4-1}
\bibliography{thermo}

\appendix

\setcounter{equation}{0}
\setcounter{figure}{0}
\setcounter{page}{1}
\renewcommand{\theequation}{S\arabic{equation}}
\renewcommand{\thefigure}{S\arabic{figure}}

\section{\large Supplementary Informations}

\section{Sufficient conditions for the strong violation of the detailed balance}

It is interesting to discuss which are the conditions to observe the strong violation of the detailed balance reported in the main text for the golden rule rate of the Eq.~(1).  For a Coulomb island, at equilibrium in the temperatures, the rates usually satisfy the detail balance
\begin{equation}
\Gamma(-\delta U)=e^{-\beta \delta U}\Gamma(\delta U),    
\end{equation}
with the inverse temperature $\beta=1/k_B T$. This property 
guarantees that the final distribution of the quantum dot states follows the standard Boltzmann distribution, such that the $n$ charge states of the island has an occupation probability $p(n)\propto e^{-\beta E_c(n)}$, with $E_c(n)$ the energy of the state \cite{nazarovbook2009}. 
Clearly however, when the island and the leads have a different temperatures $\beta_j\neq \beta_{is}$, the detail balance is necessarily violated.

In the main text, we showed a case where the violation is extreme: when the "backward" rate (that promotes electron tunneling in energetically unfavourable state) is favoured in comparison to the "forward" one, i.e. 
\begin{equation}
\label{Eq:StrongViolation}
    \Gamma(-\delta U)>\Gamma(\delta U),
\end{equation}
with $\delta U>0$. 
By noticing that the functions determining the rate are all positive, $0\leq f_j,n_j$, and the Fermi distribution satisfies $f_l\leq 1$, and by assuming the EH symmetry for the leads, one can easily see that Eq.~(\ref{Eq:StrongViolation}) can be satisfied  \emph{only if} for some energy $\delta U>0$ we have
\begin{equation}
\label{eq:inequality}
n_{is}(\epsilon-\delta U)[1-f_{is}(\epsilon-\delta U)]>n_{is}(\epsilon+\delta U)[1-f_{is}(\epsilon+\delta U)].
\end{equation}
Since this is a necessary condition it is instructive to investigate what are the requirements that the island DOS has to satisfy to promote the strong violation.  We see immediately that for energy independent DOS of the island $n_{is}(\epsilon)=const$ (non superconducting metallic island) the inequality cannot be satisfied since the Fermi function is a monotonously decreasing function of the energy, i.e. $f_j(E)\geq f_j(E')$ for $E\leq E'$. Without any other assumption on the DOS it immediately follows that an $NIN$ junction cannot ever show any strong violation of the detail balance. This argument also implies that an $SIN$ junction cannot shown any strong violation of the detail balance.

In turn, for an $SIS$ junction the inequality Eq.~(\ref{Eq:StrongViolation}) can be potentially satisfied due to the fact that, for a standard BCS DOS, it can happen that $n_{is}(\epsilon-\delta U)>n_{is}(\epsilon+\delta U)$ for $\epsilon >\Delta_{is}$, compensating the natural opposite behaviour induced by the $1-f$ term to not violate the inequality. This demonstrates the importance of the energy dependence of the DOS for, at least, some energies.  

However, one needs to exclude the case of two identical superconductors. Indeed, for a pure $SIS$ case, i.e. $n_{is}(\epsilon)= n_j(\epsilon)$, an exchange in the roles of the DOS between the lead and island is fully equivalent  to a change in sign of the temperature difference, with a consequent temperature-independent violation of the detailed balance, that implies  the violation of the second principle of thermodynamics. Thus, the condition of strong violation of the detail balance identifies a physical mechanism in which for some states in a certain range of energies it is convenient to go against the electrostatic energy.

It is important to stress again that inequality Eq.~(\ref{eq:inequality}) is a \emph{necessary} condition but \emph{not a sufficient} one to realize the reported strong violation (notice that the inequality Eq.~(\ref{eq:inequality}) depends only on the island temperature and obviously at equilibrium the detailed balance must be satisfied, implying the necessary character of Eq.~(\ref{eq:inequality}) but not the sufficient one). As discussed also in Ref. \cite{marchegiani2020,marchegiani2020b} the temperature difference has to be operated in a way that the large gap is kept at a temperature higher than the small gap. Furthermore, the condition of the monotonously  decreasing density of states for $\epsilon>0$ does not necessarily imply superconductivity as requirement. This was discussed in the Supplementary material of Ref.~\cite{marchegiani2020}, for example a Lorentzian density of state $n_{is}(\epsilon)\propto \gamma /(\gamma^2+\epsilon^2)$, if always centered in the chemical potential, can also determine the violation of the inequality, thus enlarging the range of cases where the strong violation of the detail balance can be indeed similarly reported beyond BCS superconductivity.

Finally, we wish to importantly note another requirement that needs to be satisfied in order to make possible the strong violation of the detail balance. By inspection of  Eq.~(\ref{eq:inequality}) it follows that the argument of the island DOS must shift with the energy $\delta U$, i.e. $n_{is}(\epsilon\pm \delta U)$. Indeed, the DOS shifts with $\delta U$ due to the fact that the SC gap opens at the Fermi energy as a result of the \emph{intrinsic} interacting nature of the Cooper instability.  Notably, this complexity is elegantly accounted for by the BCS mean-field theory, whose simplicity in turn may hide its profound origin. For example, in a semiconductor featuring a  similar band-gap in the spectrum,  the position of the chemical potential is determined by the electron density and it is not fixed to reside in the gap. It follows that the DOS does not shift with the electrostatic energy. In turn, in a superconductor when the electrochemical potential is changed, such as by applying a bias to a junction, the DOS shifts with it.  In the Coulomb island we are considering, the role of the bias is essentially played by the electrostatic energy difference $\delta U$. 

The importance of the interaction was never particularly stressed before but the rate analysis discussed here shows it in a crystal clear way and it also sheds light on the physics of the bipolar thermoelectricity in the whole sense \cite{marchegiani2020,marchegiani2020b}. Also in that cases we can state that the mean-field nature of the BCS theory, which properly takes in account the important part of the interaction, is a crucial ingredient behind the bipolar thermoelectricity in the tunnelling limit. This conclusion can be concurrently obtained also by the experimental observation that bipolar thermoelectricity is connected to a spontaneous breaking of the EH symmetry. In the experimental demonstration of the bipolar thermoelectric effect we have demonstrated also a current controlled hysteretic cycle of the junction, which clearly showed the reality of the spontaneous symmetry breaking of the EH symmetry.

The analysis so far presented assumed a finite $\delta U$, that also implies nonlinearity in the electrostatic energy difference. This is meaningful, since the maximum of violation of the detail balance appears at the matching-peak conditions $\delta U=\pm (\Delta(T)-\Delta_{is}(T))$. However, it is interesting to look at what happens at very low energies $\delta U\to 0$. By expanding the rate we find
\begin{widetext}
\begin{equation}
\frac{\partial\Gamma_j}{\partial \delta U}\Bigl|_{\delta U=0}= \frac{1}{e^2 R_j} \int_{-\infty}^\infty d\epsilon \left(\frac{\partial n_{is}(\epsilon)}{\partial \epsilon}[1-f_{is}(\epsilon)]-n_{is}(\epsilon)\frac{\partial f_{is}(\epsilon)}{\partial \epsilon}\right)
    n_j(\epsilon) f_j(\epsilon).
\end{equation} 
\end{widetext}
Clearly the first term is not present if the DOS of the island $n_{is}$ is not affected by the energy $\delta U$, and we are left with only the second term that is always positive. So \emph{only if} the island DOS is monotonously decreasing, i.e. $\partial_\epsilon n_{is}<0$ for $\epsilon >0$ one could even have a negative $\partial\Gamma_j/\partial \delta U<0$ which strongly violate the detail balance. So the arguments we did before in nonlinear cases ($\delta U\neq 0$) can be repeated in the linear case. Notably, the presence of a negative  $\partial\Gamma_j/\partial \delta U$ in the linear regime of $\delta U$ is necessarily associated to the interaction effects included by the island DOS by the BCS theory. A similar phenomenology is also observed at the level of the current of the SIS' junction, where at low bias a negative differential conductance is reported \cite{marchegiani2020}. This is again connected to the SSB of the EH symmetry which create an unstable situation, essentially shown in the strong violation of the detail balance, where the system tries to push particles against the natural direction dictated by the electrostatic energy. This intrinsic instability constitutes at the same time an arbitrary strong violation of the non-equilibrium fluctuation dissipation theorem, which is nevertheless valid at the temperature equilibrium due to the detail balance.

\begin{figure*}[t!]
  \includegraphics[width=2\columnwidth,trim=0cm 6cm 0cm 8cm]{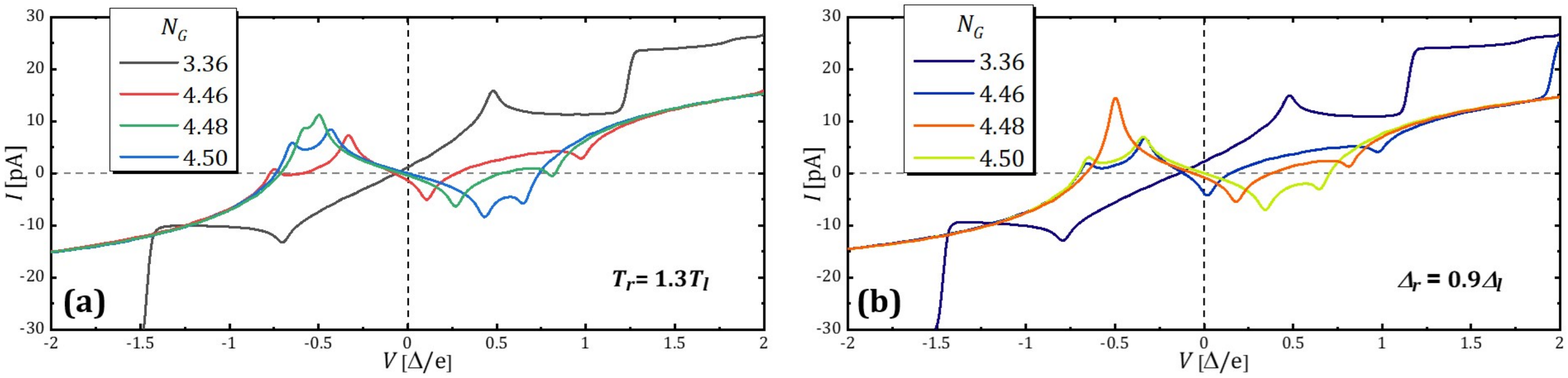}
    \caption{\textbf{(a)}:Source grain current $I$ as a function of source drain bias $V$ with $T_{cold}=T_{is}=0.2 T_C$, $T_l=0.7 T_C$ and $T_r=1.3 T_l$, for different values of $N_G$.\textbf{(b)}:Source grain current $I$ as a function of source drain bias $V$ with $\Delta_{is}=\Delta/2$, $\Delta_l=\Delta$ and $\Delta_r=0.9 \Delta$, for different values of $N_G$.}
  \label{supp1}
\end{figure*}

\section{Differences with the conventional thermoelectricity in quantum dots}

In this section we want to discuss the crucial differences between the bipolar thermoelectric effects of the system studied in the main text and the unipolar thermoelectricity that usually appears in conventional quantum dots \cite{staring1993,mahan1996}. Quantum dots can be tuned via an external gating to be \emph{non} EH symmetric \cite{averin1991}, thus generating a thermoelectric effect. It is indeed well know that the quantum dots can show a gate dependent linear thermoelectric effect when the thermal gradient is applied throughout the system, i.e. when the temperatures of the source and drain leads are different. Notably, the sign of the Seebeck coefficient can be changed by appropriately operating with a gate. The gate is indeed capable to move the system from electron-like to hole-like characted. However, we term this effect unipolar since once the gating condition is fixed, a change in the thermal gradient causes a sign change in the thermoelectrical response. Furthermore, the presence of a thermal gradient between the leads makes, in general, the structure also non-symmetric by inversion, so in case of an unipolar thermoelectric effect a non-reciprocal IV characteristic $I(V,\Delta T)\neq -I(-V,\Delta T)$ also appears.
This situation strongly differs from the \emph{bipolar} thermoelectric effect reported in main text, where the temperature difference between isalnd and leads does not determines the sign of the thermoelectricity that it is only determined by the spontaneous symmetry breaking. This is a real crucial difference between the two effects. Indeed, in the system we discuss in the main text, it is crucial that we apply a thermal gradient between the central island an the two external leads (kept at the same temperature). This configuration is not expected to generate any net thermoelectric effect classifiable as a standard \emph{unipolar} thermoelectrical contribution. Indeed,  the thermal gradients at the two junctions have opposite signs with respect to the current flow trough the system and, for an identical left/right tunnel barriers we consider, no net unipolar thermoelectrical contribution is expected to be generated. Notice that this thermal configuration is well known to not generate a net (unipolar) thermoelectric effect. As a comparison, thermoelectric cells usually combine $p$- and $n$-like material with opposite signs of the thermoelectrical coefficient (thermo-couple configuration). In conclusion, the temperature configuration considered in the system uniquely proves the bipolar nature of the effect. This has been discussed also elsewhere \cite{marchegiani2020b} but, in the main text, we demonstrated that the same arguments substantially apply also in the presence of Coulombic effects. 

At the same time, one can speculate that if the right/left barrier symmetry is broken a net linear thermoelectrical effect may eventually emerge. General considerations coming from EH symmetry of the system \cite{benenti2017}, at the linear level, suggest that this configuration would not generate any linear thermoelectric effect. However, the bipolar effect is a nonlinear effect, requiring a finite temperature difference and sometimes also finite bias, so those general linear arguments do not applied. It is interesting to explore how a different set of conditions for the system (and the system parameters) affects the reported evidences. We tested a different set of cases changing the structure in Fig. 1(a) keeping the same configuration of temperatures (cold island and with the hot leads at the same temperatures). In particular we tested all the following configurations: 
\begin{itemize}
    \item Full metallic structure (no superconductors) with equal tunnel barriers ($R_l = R_r$),
     \item Full metallic structure with different tunnel barriers ($R_l \neq R_r$),
      \item Metallic dot with superconductive leads with equal tunnel barriers ($R_l = R_r$),
      \item Metallic dot with superconductive leads with different tunnel barriers ($R_l \neq R_r$),
      \item Superconductive dot with metallic leads with equal tunnel barriers ($R_l = R_r$),
      \item Superconductive dot with metallic leads with different tunnel barriers ($R_l \neq R_r$).
\end{itemize}
\emph{None} of these configurations showed any kind of thermoelectricity. This result can be seen as the consequence of the fact that thermoelectricity in our system is strongly correlated with the nature of the BCS DOS of island and leads, as pointed out in the previous discussion on the necessary conditions for the bipolar thermoelectricity. 

\section{Robustness of the effect}
However, clearly, if the temperature of the external right/left superconducting leads are different one can expect a contribution coming for the standard unipolar thermoelectrical effect. So even in the case of not perfect temperature profile an experiment can address the signature of the reported effect. Similarly one could speculate what happen in the case that the gap of the superconducting leads are not exactly identical, as we simply assumed in the main text. In other words, hereafter we want to discuss the robustness of the effect with respect to a change in the temperatures or gaps profiles assumed, for simplicity in the main text. We will consider separately the two most interesting cases: {\it i)} the temperature of the leads is not perfectly equal and {\it ii)} the SC gap of the leads is not perfectly equal. 

Taking into analysis the first case, in Fig.~\ref{supp1}(a) it can be observed the IV characteristic when the temperature of the right lead is higher than the one of the left lead. Here we have $T_{is}=0.2 T_C$, $T_l=0.7 T_C$, $T_r=0.8 T_C$ where $T_{is}$ is the temperature of the island and $T_l$, $T_r$ are the temperatures of the left and right lead respectively. These curves are still reminiscent of the bipolar thermoelectric effect discussed in the main text. However, on top of that we can recognize the contribution of the \emph{unipolar} thermoelectric effect that manly shifts the intercept of the curves al zero bias. Indeed these IV curves loose the reciprocity, $I(V,\Delta T)= -I(-V,\Delta T)$, as a direct consequence of the emerging unipolar thermoelectricity that not symmetric temperature configuration of the external leads will determine being associated to the \emph{conventional} thermoelectrical behavior of the quantum dots. However, if the quantum dot is gated exactly at the Coulomb resonance, such as $N_G=4.5$, the EH symmetry of the symmetry is again restored: no linear unipolar thermoelectric effects and the reciprocity of IV curves is fully restored. However, also in that case, intriguing differences can be seen with respect to the result reported in the main text. In particular the difference between the IV curvers where two thermoelectric peaks appear at the Coulomb resonance where in the main text we have only a one-peak matching peak for that gate value. Indeed the temperature difference of the leads determine also a different value for the superconducting gaps in the leads, i.e. $\Delta_r(T_r)\neq \Delta_l(T_l)$, as predicted by BCS \cite{tinkham2004}. So, correspondely, there are two different matching peaks values depending on which barriers the tunnelling happens. This result in the double peak structure also at resonance. However, even if some of the peculiarity of the bipolar thermoelectric effect are lost, at least comparing Fig.~\ref{supp1}(a) with the main text figures, we can still conclude the existence of \textit{bipolar} thermoelectric effect component. Many of the figure of merits discussed in the paper are roughly still valid and are not substantially ruined by a moderate temperature inequality in the leads.  

Finally we wish to show the possible consequence of the fact that two external lead have a slightly different zero temperature gaps. However, for simplicity, the temperature configuration is taken exaclty the same of the main text. In this case it can be seen from Fig.~\ref{supp1}(b) that the situation remind aspects of the previous case. Anyway, here also the right/left symmetry is broken. In this case too, we can conclude that the effect described in the main text is sufficiently robust and do not need fine tuning of the parameters to be observable.

\end{document}